# Accurate prediction of thermal conductivity of $Al_2O_3$ at ultrahigh temperatures

Janak Tiwari, Tianli Feng*

Department of Mechanical Engineering, University of Utah, Salt Lake City, UT 84112, USA

**Corresponding Author:**

*tianli.feng@utah.edu

ABSTRACT

Many complex crystals show a flattening or even increasing lattice thermal conductivity at high temperatures, which deviates from the traditional 1/T decay trend given by conventional phonon theory. In this work, we predict the thermal conductivity of $Al_2O_3$ that matches with experimental data from room temperature to near melting point (2200 K). The lattice thermal conductivity is found to be composed of contributions of phonon, diffuson, and radiation. Phonon particle thermal conductivity decays approximately as $\sim T^{-1.14}$ after considering four-phonon scattering as well as finite temperature corrections to lattice constant, harmonic, and anharmonic force constants. Diffuson (inter-band tunneling) thermal conductivity increases roughly as $\sim T^{0.43}$. Radiation thermal conductivity increases as $\sim T^{2.51}$, being slightly smaller than $\sim T^3$ due to the increase of phonon linewidth with temperature, which increases photon extinction coefficient and reduces photon mean free path. At room temperature, phonon, diffuson, and radiation contribute about 99%, 1%, and 0, respectively. At 2200 K, the contributions change to 61%, 20%, and 19%, respectively. Four-phonon scattering is important at ultra-high temperature, decreasing the phonon thermal conductivity by a maximum of 24%. The finite-temperature softening effects of harmonic and anharmonic force constants increase the phonon thermal conductivity by a maximum of 36% at ultra-high temperatures. We also verify that Green-Kubo MD can capture phonons' both particle and wave natures, similar to the Wigner formalism. At ultra-high temperatures, the photon mean free path is found to be in the order of 100 nm, which should be considered for experimental measurement of thin films. This study aims to enhance the understanding of lattice thermal conductivity in complex crystals at ultra-high temperatures, potentially spurring further exploration of materials suitable for such extreme conditions.

## I. INTRODUCTION

Thermal conductivity of complex crystals at high temperatures is critical for many applications, such as thermal barrier coatings [1,2], refractory materials [3,4], crucibles, and high-temperature thermal insulation [5,6] However, current state-of-the-art theoretical prediction based on phonon-gas model (PGM) [7–9] could not explain their intriguing thermal conductivity ($\kappa$). At intermediate temperatures, $\kappa$ decays with temperature (T), being consistent with typical crystal behavior. However, at high temperatures, $\kappa$ either increases or remains independent of temperature, displaying a peculiar and anomalous glass-like behavior. While the former behavior can be explained using PGM based on the Boltzmann transport equation (BTE), which primarily involves three-phonon (3ph) or four-phonon (4ph) scattering processes, the latter phenomenon remains a puzzle, presenting an unanswered question.

In this study, we take $Al_2O_3$ as an example to investigate the flatting or increasing trend of thermal conductivity at high temperatures for the following reasons. (1) $Al_2O_3$ has excellent mechanical strength, high-temperature thermal and chemical stability [10–12], large band gap [13], high dielectric constant [14], high melting point [15], and is widely used in many high-temperature engineering applications. (2) It does not have an electronic contribution to heat transfer, leaving the theoretical lattice thermal conductivity readily comparable to experimental data. (3) It can be grown to large-size single crystal with high purity, so grain boundary and defect scattering can be neglected. (4) Extensive experimental data [16–25] are available to validate our study.

This study looks into several possible improvements to the current state-of-the-art $\kappa$ prediction to explain the flatting or increasing trend of thermal conductivity in complex crystals at ultra-high temperatures. One improvement is the temperature correction on lattice structure and phonon-phonon scattering. The current phonon theory relies on the ground state lattice structure and force constant, where 3ph $\propto$ phonon population (n) $\propto$ T and 4ph $\propto n^2 \propto T^2$ [26,27], resulting to phonon thermal conductivity ($\kappa_{ph}$) decay with power law $T^{-1}$ and $T^{-2}$ respectively for 3ph and 4ph scattering. However, phonon renormalization at high temperatures due to lattice expansion and temperature dependence of harmonic force constant changes 3ph and 4ph scattering phase space [28], which may lead to an increase in the $\kappa_{ph}$. Recently, it has been reported that temperature dependence of anharmonic force constant reduces the scattering probability (or scattering cross-section) [28,29], which increases $\kappa_{ph}$. These make the thermal conductivity flatter at higher temperatures compared to ground state calculations.

The second improvement could be the incorporation of the diffuson thermal conductivity ($\kappa_{dif}$ or $\kappa_c$) along with $\kappa_{ph}$. In PGM and BTE, the primary heat carriers are propagating phonons (particle-like phonons), which account only for the diagonal terms of the velocity operator. This is a good approximation for simple crystals [30], where the phonon branches are well separate, i.e., inter-band spacings are much larger than the phonon linewidths. However, this approximation fails in the complex crystals and disordered regime [31], where phonon bands are not well separated and many phonon modes might overlap with each other. In this scenario, phonon shows wavelike transport properties i.e., they can tunnel from one mode into another and conduct heat. Recently, Simoncelli *et. al.* [32,33] introduced the Wigner transport equation which encompasses particle-like and wavelike conduction mechanisms, providing a unified approach to heat-transport phenomena in solids, including simple crystals (where particle-like propagation dominates), glasses (where wavelike tunneling or diffusons dominates), and all intermediate cases (complex crystals where both particle-like and wavelike conduction mechanisms coexist). The significant $\kappa_{dif}$ is reported for many complex crystals [34–37] and used to explain the flattening of $\kappa$ at higher temperatures.

The temperature effect on both $\kappa_{ph}$ and $\kappa_{dif}$ could also be studied using molecular dynamics (MD) simulations. In principle, MD should capture all orders of phonon-phonon interaction as well as diffuson, due to its ability to simulate the intricate behavior of individual atoms, thereby capturing all the underlying microscopic mechanisms governing lattice heat transfer. MD, while powerful, is limited by the accuracy of the interatomic potential functions they rely on. Classical potentials, based on empirical models, often involve approximations and force field parameters, leading to inherent inaccuracies in capturing the complex quantum mechanical behaviors of atoms and molecules. Machine learning interatomic potentials (MLIPs), on the other hand, have very high accuracy, comparable to that of density functional theory (DFT) calculations. In this study, we train moment tensor potential [38–40] (MTP) from *ab initio* molecular dynamics (AIMD) snapshots and used it to run Green-Kubo molecular dynamic (GKMD) [41,42] to calculate thermal conductivity ($\kappa_{GKMD}$).

The third improvement could be the incorporation of radiation thermal conduction ($\kappa_{rad}$) to overall thermal conductivity. Similar to phonon, photon also propagates inside and through the crystal and transports the heat, which contributes to apparent thermal conduction. Radiation contribution has been hypothesized as a primary reason for the increase in thermal conductivity at ultra-high temperatures in early literature [18,21,24,43–46]. However, the radiation thermal conductivity has not been calculated rigorously from first principles. Here, we calculate the radiative properties using the Lorentz oscillator model [47,48] and estimate the radiation thermal conductivity based on the Rosseland model.

This study presents the thermal transport of $Al_2O_3$ from room temperature to melting point (2200K) by incorporating the temperature-dependent force constants in $\kappa_{ph}$ calculation and considering the $\kappa_{dif}$ and $\kappa_{rad}$. $\kappa_{ph}$ and $\kappa_{dif}$ contributions are calculated using Wigner formalism and Green-Kubo molecular dynamics (GKMD) separately. The remaining section of the paper is structured as follows. In section II, we present the computations details and methodology used in the study. Section III and IV presents the main results and discussions respectively. Finally, section V presents the conclusions.

## II. METHODOLOGY

### A. Computational workflow

Figure 1 shows the computational workflow of the study. First of all, the structure is relaxed by using the Vienna Ab initio Simulation Package (VASP) [49,50], using generalized gradient approximational (GGA) [51] method and Perdew-Burke-Ernzerhof revised for solids (PBEsol) functional [52]. The plane-energy cutoff used in the calculations is 500 eV, and the energy and force convergence thresholds are $1\times10^{-8}$ eV and $1\times10^{-7}$ eV·Å$^{-1}$, respectively. Al$_2$O$_3$ belongs to the trigonal lattice structure system (space group $R\overline{3}c$), with a rhombohedral primitive unit cell containing 10 atoms. The relaxed lattice parameters are $a$ = 5.139 Å and α = 55.35º, which closely resembles the experimental value [53,54] of $a$ = 5.128 Å and α = 55.28º. In DFT calculations, supercells of 3×3×3 (270 atoms) are used with a **k**-point grid of 4×4×4. Other parameters are kept constant as that of relaxation.

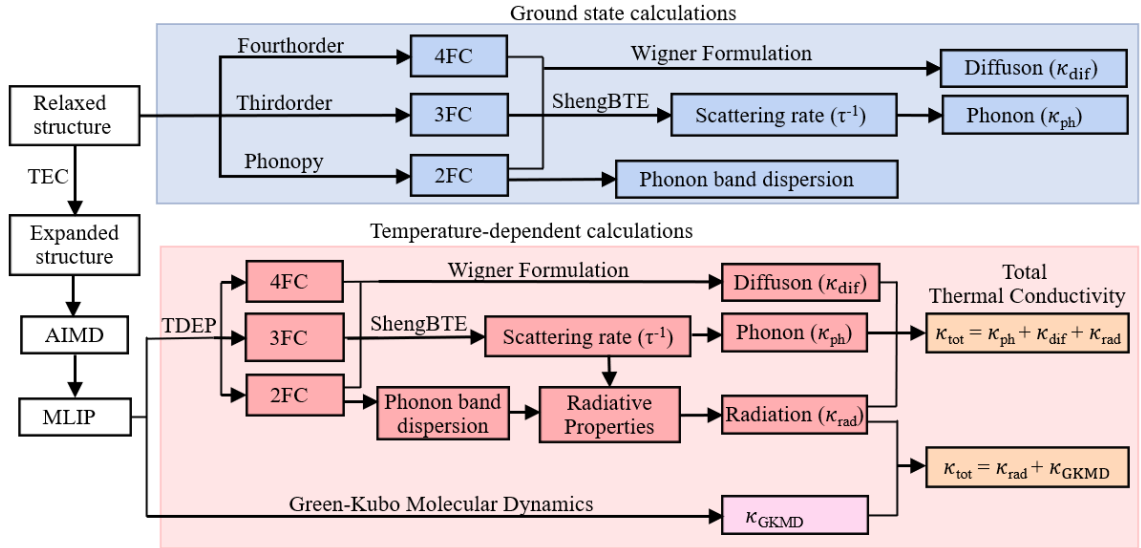

*Figure 1: Computation workflow of the study.*

For the ground state calculations, harmonic or 2$^{nd}$-order force constants (2FC) are extracted using Phonopy [55] considering the 4$^{th}$ nearest neighbors. The anharmonic force constants (AFC), including 3$^{rd}$ (3FC) and 4$^{th}$ order (4FC), are calculated using Thirdorder and Fourthorder packages built inside ShengBTE [56], considering the 4$^{th}$ and 2$^{nd}$ nearest atoms, respectively. For the finite temperature calculations, the structure is expanded using thermal expansion coefficient (TEC). The *ab initio* molecular dynamics (AIMD) is computed and snapshots are recorded, from which moment tensor potential (MTP) [39,40,57] based machine learning interatomic potential (MLIP) is trained. GKMD is performed by using the MLIP-LAMMPS package [39,58] to calculate $\kappa_{GKMD}$. The temperature-dependent harmonic and anharmonic force constants are extracted using the temperature-dependent effective potential (TDEP) method [59,60] and MPT. Using the force constants, the $\kappa_{ph}$ and $\kappa_{dif}$ are calculated by solving Wigner's formalism [32,33] under ShengBTE. Radiative properties are calculated using the Lorentz oscillator model, from which $\kappa_{rad}$ is calculated using the Rooseland model. Finally, the total thermal conductivity ($\kappa_{tot}$ or $\kappa$) is calculated by summing up $\kappa_{ph}$, $\kappa_{dif}$, and $\kappa_{rad}$ as well as $\kappa_{GKMD}$ and $\kappa_{rad}$ separately. The details of each step are explained below.

**B. Thermal Expansion Coefficient**

The linear thermal expansion coefficient (TEC) is calculated using quasi-harmonic approximation (QHA) with the following formalism [61]:

$$\alpha_L = -\frac{k_B}{3N_\mathbf{q} V_c B} \sum_{\mathbf{q},j} \gamma_{\mathbf{q},j} \cdot \left(\frac{x}{2}\right)^2 \cdot \left[1 - \coth^2\left(\frac{x}{2}\right)\right].$$

Here, $(\mathbf{q}, j)$ stands for a phonon mode with wavevector $\mathbf{q}$ and dispersion branch $j$. $N_\mathbf{q}$ is the number of phonon $\mathbf{q}$ points. $V_c$ is the volume of a primitive cell of $Al_2O_3$. $k_B$ is the Boltzmann constant. $B = -V\frac{dP}{dV}$ is the bulk modulus. The summation is done over all the $3N_\mathbf{q} n_b$ phonon modes, where $n_b$ is the number of basis atoms in a primitive cell. $x$ is short for $x = \hbar\omega_{\mathbf{q},j}/k_B T$. $\gamma_{\mathbf{q},j} = -\frac{V}{\omega_{\mathbf{q},j}}\frac{\partial \omega_{\mathbf{q},j}}{\partial V}$ is the mode-dependent Grüneisen parameter. In this study, an 18×18×18 $\mathbf{q}$-mesh is used in the TEC calculations.

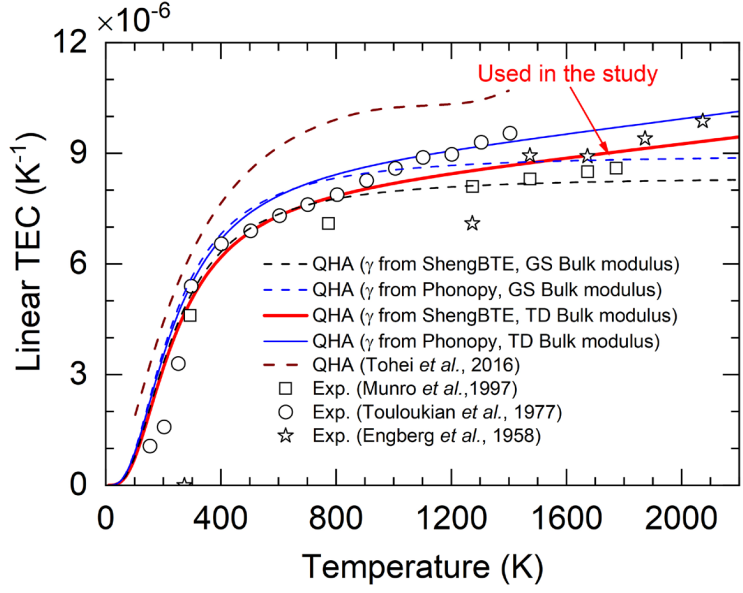

*Figure 2: Linear thermal expansion coefficient of Al₂O₃ as a function of temperature. The theoretical TEC (lines) are calculated by quasi-harmonic approximation (QHA) using the Grüneisen parameter and bulk modulus obtained from different ways. The ground state (GS) bulk modulus is calculated by DFT, and temperature-dependent (TD) bulk modulus is taken from Ref. [17]. The DFT predicted data by Tohei et al. [62] and different experimental data [17,63,64] are included for comparison.*

The bulk modulus at the ground state is calculated using VASP by finite difference method using $B = -V\frac{dP}{dV}$, which gives $B$ = 249 GPa. It matches well with the experimental data of 248.7 GPa [65] and 257 GPa [17]. The Grüneisen parameters are obtained by two methods. One is using finite difference method, i.e., $\gamma_{\mathbf{q},j} = -\frac{V}{\omega_{\mathbf{q},j}}\frac{\partial \omega_{\mathbf{q},j}}{\partial V}$, implemented in Phonopy [55]. The other method is to use the third-order anharmonic force constant to predict $\gamma_\lambda$ as implemented in ShengBTE [56]. As shown in Fig. 2, the predicted TEC using the ground state bulk modulus (blue-dash curve with Grüneisen parameter from Phonopy and black-dash curves with Grüneisen parameter from ShengBTE) match experimental data at low to medium temperatures but deviates at ultra-high temperatures, which is commonly seen for QHA. This deviation has been reported to be corrected by using the temperature-dependent bulk modulus in our previous work [28]. With the temperature-dependent bulk modulus [17], and the predicted TEC agrees with experimental data even at ultra-high temperatures [17,63,64].

## C. Machine Learning Interatomic Potential

We employ a moment tensor potential (MTP) [39,40,57] based machine learning interatomic potential (MLIP) to characterize the temperature-dependent potential surface of $Al_2O_3$. The accuracy and effectiveness of this method have been demonstrated in previous studies [29,66]. A potential is trained for each temperature of study, i.e., at 500, 800, 1000, 1200, 1500, 1600, 1700, 1800, 1900, 2000, 2100, and 2200 K. The temperature interval is kept small at higher temperatures as the focus of the study is to study the flattening trend of $\kappa$ at higher temperatures. The training database for each potential is prepared by AIMD with NVT ensemble for 500 steps with time step of 5 fs. The lattice structure for particular temperatures is expanded using TEC. A supercell of 3×3×3 primitive cell containing 270 atoms is used in the simulation domain. Four independent AIMDs with randomly displaced initial atomic positions are performed at each temperature to better sample the potential energy surface. Energies, forces, and stresses are recorded together with corresponding atomic configurations to construct the training and testing database. The database is separated randomly maintaining 75% (1500 snapshots) for training and 25% (500 snapshots) for testing. The initial MTP of level 22 is selected to train potential based on the accuracy and computational demand. The selected initial potential is trained for 1000 iteration steps with the minimum and maximum atomic interaction cutoff of 1.2 Å and 5.5 Å, respectively.

**D. Green-Kubo molecular dynamics**

Once the MTP with a small error is developed, GKMD is performed by using the MLIP-LAMMPS package [39,58]. GKMD calculates the lattice thermal conductivity by integrating the heat current autocorrelation function based on the Green–Kubo formula [41,42]:

$$\kappa_{GKMD} = \frac{1}{3k_B T^2 V} \int_0^\infty \langle \vec{J}(0) \cdot \vec{J}(t) \rangle \, dt \tag{1}$$

where $k_B$ is the Boltzmann constant, $V$ is volume of total simulation domain, $T$ is temperature, $\vec{J}(t)$ is the heat current, and the angular bracket represents an autocorrelation. In LAMMPS, the heat current vector is calculated by the energy and forces of the system, which is obtained from the MTP.

In this study, we employ a 7×4×3 supercell of the conventional cell, containing 5042 atoms. The size effect is studied. Periodic boundary conditions are implemented in all three dimensions. The time step of GKMD is set to 1 fs. First, an NVT ensemble is run for 200,000 steps (0.2 ns) to fully stabilize the temperature of the system. Then, an NVE ensemble is run for 200,000 steps (0.2 ns) to fully stabilize the system. Finally, another NVE ensemble is run for 800,000 steps with a correlation time of 200 ps, during which the heat current correlation is recorded. To mitigate the noise and intrinsic statistical error of GKMD, we conduct 16 independent runs with different initial velocities for each temperature. The ratio between the total running time and the correlation time is maintained larger than 300, which has been reported to be sufficient [67]. Since the temperatures in this study are relatively high, the difference between a classical statistic and a quantum statistic is neglected.

**E. Temperature correction to force constants**

The temperature-dependent harmonic and anharmonic force constants are extracted using the temperature-dependent effective potential (TDEP) method [59,60] at 500, 1000, 1500, and 2000K. The TDEP method extracts effective force constants at a certain temperature by fitting the potential energy of a series of atomic trajectory images at that temperature to the 2$^{nd}$, 3$^{rd}$, and 4$^{th}$ orders. 1000 randomly displaced configurations are generated at each temperature to sample the potential surface. The forces, stress, and energies of these configurations are obtained from MTP at each temperature. The effect of the temperature is factored in by a thermal expansion, a temperature-dependent MTP trained from AIMD simulations at different temperatures, and a temperature-dependent displacement of atoms in the generated supercells.

**F. Phonon and diffuson thermal conductivity by Wigner formalism**

Using the force constants, the thermal conductivity is calculated by solving the Wigner's formalism [32,33]:

$$\kappa^{\alpha\beta} = \kappa_{ph}^{\alpha\beta} + \kappa_{dif}^{\alpha\beta}, \tag{2}$$

$$\kappa_{ph}^{\alpha\beta} = \frac{\hbar^2}{k_B T^2 V_c N_\mathbf{q}} \sum_\mathbf{q}^{N_\mathbf{q}} \sum_j^{3n_b} v_{\mathbf{q},j}^\alpha v_{\mathbf{q},j}^\beta \omega_{\mathbf{q},j}^2 n_{\mathbf{q},j}(n_{\mathbf{q},j} + 1)\tau_{\mathbf{q},j}, \tag{3}$$

$$\kappa_{dif}^{\alpha\beta} = \frac{\hbar^2}{k_B T^2 V N_\mathbf{q}} \sum_\mathbf{q}^{N_\mathbf{q}} \sum_{j \neq j'}^{3n,3n} v_{\mathbf{q},jj'}^\alpha v_{\mathbf{q},j'j}^\beta \frac{\omega_{\mathbf{q},j} + \omega_{\mathbf{q},j'}}{2} \frac{\omega_{\mathbf{q},j} n_{\mathbf{q},j}(n_{\mathbf{q},j} + 1) + \omega_{\mathbf{q},j'} n_{\mathbf{q},j'}(n_{\mathbf{q},j'} + 1)}{4(\omega_{\mathbf{q},j'} - \omega_{\mathbf{q},j})^2 + (\tau_{\mathbf{q},j}^{-1} + \tau_{\mathbf{q},j'}^{-1})^2} (\tau_{\mathbf{q},j}^{-1} + \tau_{\mathbf{q},j'}^{-1}), \tag{4}$$

$$\tau_{\mathbf{q},j}^{-1} = \tau_{3ph}^{-1} + \tau_{4ph}^{-1} + \tau_{ph-iso}^{-1}. \tag{5}$$

Here $\alpha$ and $\beta$ are cartesian directions. $\kappa_{ph}^{\alpha\beta}$ is the phonon particle thermal conductivity (or Peierls thermal conductivity), obtained from ShengBTE. $\kappa_{dif}^{\alpha\beta}$ is the diffuson thermal conductivity, calculated using an in-house code. $j$ and $j'$ are phonon branches, $\omega$ is the angular frequency, $v$ is the group velocity, and $\tau^{-1}$ is the phonon scattering rates, including three-phonon, four-phonon, and phonon-isotope scatterings. The convergence of ShengBTE at various **q**-mesh densities is tested. Specifically, the 3ph calculation converges at a **q**-mesh density of 18×18×18, and the 3ph + 4ph calculation converges at 6×6×6. The iterative solution to three-phonon and phonon-isotope scattering is included, as implemented in ShengBTE. Four-phonon scattering is taken at the relaxation time approximation (RTA) level. The formalism of 3ph and 4ph scattering rates can be found in Ref. [27].

**G. Radiation thermal conductivity**

Similar to phonon, photon can also transport in materials. Analogous to phonon creation and annihilation (i.e., phonon scattering events), which limit phonon mean free path, photon absorption and re-emission also limit photon mean free path inside a material. When a material's size is much larger than phonon mean free path, phonon transports diffusively, and the phonon thermal conductivity is approximately proportional to phonon mean free path. Similarly, when a material's size is much larger than its photon mean free path, the material is optically thick, and its radiation contribution to thermal conductivity is proportional to photon mean free path. Based on the Rosseland model, the radiation contribution to thermal conductivity ($\kappa_{rad}$) of an optically thick material [18,68,69] is:

$$\kappa_{rad} = \frac{16n^2(T) \cdot \sigma_{SB} T^3}{3\beta(T)}, \tag{6}$$

where $n(T)$ and $\beta(T)$ are the temperature-dependent refractive index and extinction coefficient, respectively. $\beta(T)$ gives the attenuation of the electromagnetic waves inside the material and is the inverse of photon mean free path. $n(T)$ and $\beta(T)$ are given by

$$n(T) = \int_0^\infty n(\lambda, T) \frac{\partial E_{b\lambda}}{\partial E_b} d\lambda, \tag{7}$$

$$\frac{1}{\beta(T)} = \int_0^\infty \frac{1}{\alpha(\lambda, T)} \frac{\partial E_{b\lambda}}{\partial E_b} d\lambda. \tag{8}$$

$\sigma_{SB}$ is the Stefan-Boltzmann constant. The spectral refractive index $n(\lambda, T)$ and spectral absorption coefficient $\alpha(\lambda, T)$ can be calculated from the dielectric function $\epsilon(\lambda, T)$ as:

$$n^2(\lambda, T) = \frac{1}{2}\left(|\epsilon(\omega, T)| + \epsilon_{re}(\omega, T)\right), \tag{9}$$

$$\alpha(\lambda, T) = \frac{4\pi k}{\lambda} = \frac{2\sqrt{2}\pi}{\lambda}\left(|\epsilon(\omega, T)| - \epsilon_{re}(\omega, T)\right)^{\frac{1}{2}}, \tag{10}$$

$\epsilon_{re}$ is the real part of $\epsilon$. $k(\lambda, T)$ is the spectral extinction coefficient given by

$$k(\lambda, T) = \frac{1}{\sqrt{2}}\left(|\epsilon(\omega, T)| - \epsilon_{re}(\omega, T)\right)^{\frac{1}{2}}. \tag{11}$$

The dielectric function $\epsilon(\omega, T)$ can be predicted using the four-parameter Lorentz oscillator model [47,48] as:

$$\epsilon(\omega, T) = \epsilon(\infty) \prod_j \frac{\omega_{j,LO}^2 - \omega^2 + i\Gamma_{j,LO}\omega}{\omega_{j,TO}^2 - \omega^2 + i\Gamma_{j,TO}\omega}, \tag{12}$$

where $\Gamma$ is the same as phonon scattering rate, i.e., $\Gamma = \tau^{-1}$. $\omega$ is the phonon or photon angular frequency, i.e., $\omega = 2\pi f$. $j$ runs through all the infrared active transverse optical (TO) and longitudinal optical (LO) branches. $i$ is the imaginary unit number. $\epsilon(\infty)$ is the dielectric function at the high-frequency limit which is calculated from the density-functional-perturbation-theory (DFPT) [70]. The calculated values are 3.21 and 3.23 for ordinary and extraordinary rays, which match well with the experimental values of 3.2 and 3.1 [71]. The details of the calculation of $\epsilon(\omega, T)$ is similar to that of Ref. [72].

## III. RESULTS

### A. Machine learning interatomic potential and Green-Kubo MD

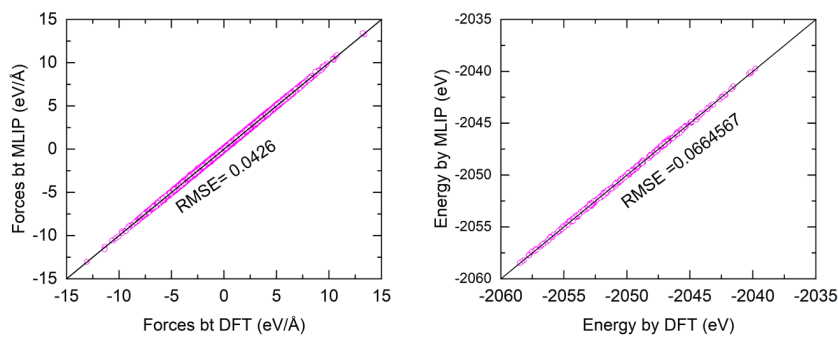

*Figure 3: Comparison of (a) forces and (b) energies obtained from machine-learned interatomic potential with DFT calculations.*

The MTP potential is trained by using the stress, forces, and energies obtained from AIMD calculations. The training and testing errors are less than 5% for each potential trained at various temperatures. To further verify the accuracy of the trained potentials, we extracted the forces and energies of the test dataset using MTP potential. Note that, this test dataset is not used to train the potential and is selected randomly. The extracted forces and energy from MTP are plotted against the forces and energies obtained from DFT calculations. As seen in Fig. 3, all the points lie along the diagonal with low root mean square error (RMSE) of 0.0426 eV/Å and 0.0664 eV for forces and energies respectively. This shows that MTP could accurately reproduce the forces and energies of the configurations and represent the actual potential surface with accuracy comparable to DFT calculation.

Using the MTP, GKMD is conducted to calculate the lattice thermal conductivity ($\kappa_{GKMD}$) as shown in Fig. 4. The average thermal conductivity is obtained by integrating the auto-correlated heat flux with correlation time. As seen in the inset of Fig. 4, the average thermal conductivity converges, suggesting that the parameters used in our calculations are appropriate. The $\kappa_{GKMD}$ shows a flatting and slight increasing trend at ultra-high temperatures, which agrees reasonably well with experimental data. This further demonstrates the high accuracy of the trained MTP. However, the experimental data shows much increasing thermal conductivity at ultra-high temperature, which could not be captured by the GKMD.

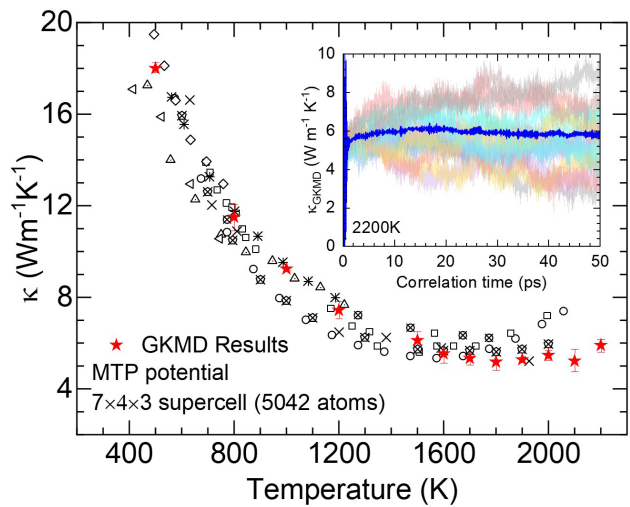

Figure 4: Thermal conductivity of $Al_2O_3$ at various temperatures obtained by GKMD using MLIPs. The experimental data (open dots) [16–25] are shown for comparison. Inset shows several independent runs and their average as a function of correlation time for 2200 K.

**B. Phonon, diffuson, and radiation thermal conductivity**

In the following, we track the changes of κ at low temperature (300 K), high temperature (1200 K), and ultra-high temperature (2200 K) when we gradually increase the calculation comprehensivity and include the contributions of phonon ($\kappa_{ph}$), diffuson ($\kappa_{dif}$), and radiation ($\kappa_{rad}$). First, we calculate the basic 3ph thermal conductivity using ground-state force constants (GSFC), shown in the blue-dash line in Fig. 5(a), which follows $\kappa \sim T^{-1}$ law. The $\kappa_{ph}$ is 28.59, 6.26, and 3.4 W·m$^{-1}$·K$^{-1}$ at 300, 1200, and 2200 K respectively. The 3ph rates at 300K and 2000K are shown in Fig. 6(a), which increases with temperature. Second, we include the 4ph scattering using GSFC, which decreases $\kappa_{ph}$, shown in dash-dot blue line, by 8%, 18% and 24% to 26.30, 5.09, and 2.48 W·m$^{-1}$·K$^{-1}$ at the three temperatures, respectively. The effect of 4ph rates is not that strong. It is primarily due to the fact that the crowd phonon branches (Fig. 6(b)) in Al$_2$O$_3$ allow the energy and momentum selection rules of three-phonon scattering easy to be satisfied. With strong three-phonon scattering, the relative importance of four-phonon scattering is naturally small. Third, we replace GSFC by temperature-dependent FC (TDFC), which is found to increase $\kappa_{ph}$ by 8%, 13%, and 36% to 28.43, 5.73, 3.37 W·m$^{-1}$·K$^{-1}$ at the three temperatures, respectively. This increase is due to the combined effect of temperature correction to the lattice constant and interatomic force constants. At elevated temperatures, phonon dispersion gets softened due to lattice expansion and harmonic force constants softening (Fig. 6(b)). This results in an increase in 3ph and 4ph scattering phase space as shown in Fig. 6(e) and 6(f), due to the change in energy and momentum conservation selection rule. At the same time, finite-temperature anharmonic force constants softening decreases the scattering cross-section as shown in Fig. 6(c) and (d), which results in the decrease in 3ph and 4ph rates as shown in Fig. 6(g) and (h). In fact, TD 2FC tends to increase the scattering rates by increasing the scattering phase space. It is due to the more pronounced reduction in TD 3FC and TD 4FC that leads to the decrease in scattering rates. Overall, incorporating TDFC increases $\kappa_{ph}$ and flattens $\kappa_{ph}$ curve at high temperatures.

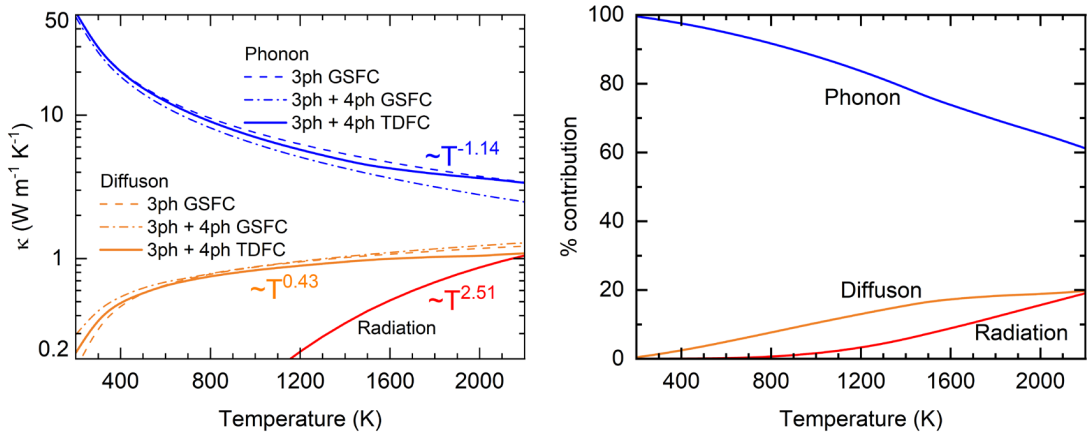

*Figure 5: (a) Phonon, diffuson, and radiation thermal conductivities in $Al_2O_3$ as a function of temperature. The power laws are fitted using the calculated data. (b)The relative contributions from phonon, diffuson, and radiation.*

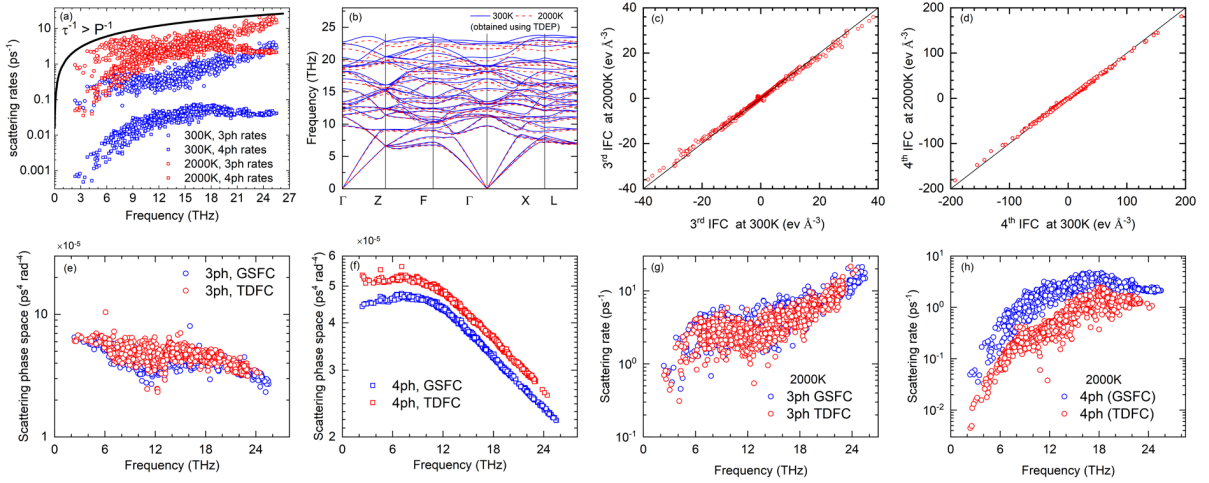

*Figure 6: (a) 3ph and 4ph scattering rates at 300 and 2000 K, respectively. The black line serves as a guideline for the comparison of phonon mean lifetime with its period. (b) Temperature dependence of phonon dispersion (calculated at 300 and 2000K using the force constant obtained from TDEP). (c,d) The temperature softening of 3$^{rd}$ IFC and 4$^{th}$ IFC. (e,f) Comparison of scattering phase space using ground state force constants and temperature-dependent force constants for 3ph and 4ph. (g,h) Decrease in 3ph rates and 4ph rates at 2000K due to temperature correction to force constants.*

One interesting phenomenon seen in Fig. 5(a) is that the $\kappa_{ph}$ using GSFC and 3ph rates (dash-blue line) and $\kappa_{ph}$ using TDFC and 3ph+4ph rates (solid-blue line) agree with each other with reasonable accuracy at high temperatures. This is due to the competing effect of 4ph scattering and TDFC in thermal conductivity, considering 4ph rates decrease $\kappa_{ph}$ while TDFC increases it. In the case of Al$_2$O$_3$, these two opposite effects are found to cancel each other. This also introduces the possibility of an error-cancellation effect on other complex materials, where the effect of TDFC and 4ph rates cancel each other, and $\kappa_{ph}$ computed using GSFC and 3ph scattering matches with experimental data.

The diffuson thermal conductivity ($\kappa_{dif}$) increases with temperature at low and intermediate temperatures and becomes flat at ultrahigh temperatures as shown in Fig. 5(a). Using GSFC, the $\kappa_{dif}$ is around 0.44 W·m$^{-1}$·K$^{-1}$ at 300K which increases to 0.95 and 1.28 W·m$^{-1}$·K$^{-1}$ at 1200 and 2200 K respectively. At high temperatures, the phonon linewidth increases, which increases the phonon tunneling probability and increases $\kappa_{dif}$. Using TDFC, the $\kappa_{dif}$ decreases by 13%, 7%, and 16% to 0.38, 0.89, and 1.08 at 300,1200 and 2200 K respectively. Adding $\kappa_{dif}$ to $\kappa_{ph}$ makes $\kappa$ flatter at high temperatures as shown in dash-orange line in Fig. 7. $\kappa_{ph} + \kappa_{dif}$ matches experimental data at intermediate to high temperatures but could not explain the increase in thermal conductivity at ultra-high temperatures.

Also, $\kappa_{ph} + \kappa_{dif}$ obtained from the Wigner formalism matches with the GKMD results throughout the temperatures. This alignment underscores the capability of both GKMD and the Wigner formalism to effectively capture $\kappa_{ph}$ and $\kappa_{dif}$. Further, the good agreement between these two different approaches supports the Wigner formalism as well as the accuracy and consistency of our calculations.

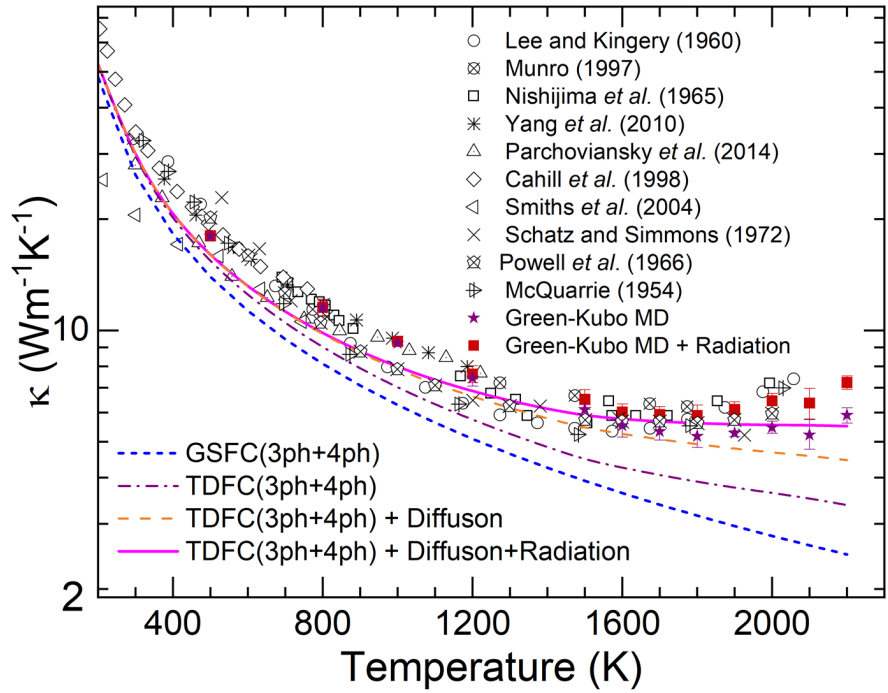

*Figure 7: The variation of thermal conductivity prediction of $Al_2O_3$ using various approaches. Experimental thermal conductivity data from Refs. [16–25] are also shown for comparison.*

The radiation thermal conductivity ($\kappa_{rad}$) is negligibly small at room temperature, but it increases with the third power of temperature and reaches 0.22 and 1.05 Wm$^{-1}$K$^{-1}$ at 1200 and 2200 K, respectively. The partial contributions of $\kappa_{ph}$, $\kappa_{dif}$, and $\kappa_{rad}$ are shown in Fig. 5(b). As seen, $\kappa_{ph}$ decays with temperature, $\kappa_{dif}$ increases and saturates with temperature, while $\kappa_{rad}$ increases rapidly with temperature. At room temperature, $\kappa_{dif}$ and $\kappa_{rad}$ are almost negligible resulting $\kappa \approx \kappa_{ph}$ i.e., $\kappa_{ph}$ contributes to almost 100% of $\kappa$. At a high temperature of 1200 K, $\kappa_{dif}$ increases significantly and contributes 13%, and the $\kappa_{rad}$ contribution is 3.27%. At an ultra-high temperature of 2200 K, $\kappa_{rad}$ becomes significant as well. At this temperature, the $\kappa_{ph}$, $\kappa_{dif}$, and $\kappa_{rad}$ contributes 61.2%, 19.7%, and 19.1% respectively.

When the $\kappa_{ph}$, $\kappa_{dif}$, and $\kappa_{rad}$ are summed up together, $\kappa$ reaches 28.81, 6.85, and 5.51 W·m$^{-1}$·K$^{-1}$ at 300, 1200, and 2200K respectively. This is shown in solid-magenta line in Fig. 6, which matches well with the experimental data. This shows that the total thermal transport comes from the contribution of $\kappa_{ph}$, $\kappa_{dif}$, and $\kappa_{rad}$. The red points in the graph show the $\kappa$ obtained by summing up $\kappa_{GKMD}$ and $\kappa_{rad}$, which also matches with experimental data.

The scaling laws of $\kappa_{ph}$, $\kappa_{dif}$, and $\kappa_{rad}$ with respect to temperature are shown in Fig. 5. For Al$_2$O$_3$, we find that $\kappa_{ph}$ decays approximately as $\sim T^{-1.14}$ after considering four-phonon scattering as well as finite temperature corrections to lattice constant, harmonic, and anharmonic force constants. This is slightly different from $\kappa_{ph} \sim T^{-1.19}$ obtained from using ground state force constants. $\kappa_{dif}$ increases roughly as $\sim T^{0.43}$. $\kappa_{rad}$ increases as $\sim T^{2.51}$, being slightly smaller than $\sim T^3$ due to the increase of phonon linewidth with temperature, which increases photon extinction coefficient and reduces photon mean free path. These scaling laws are of interest when interpreting or predicting thermal conductivity trends of other materials as well. Details of power law fittings can be found in the Supplemental Material [73].

## IV. DISCUSSION

### A. Phonon lifetime and mean free path

Based on the diffuson theory, the phonons are characterized as either normal phonons or diffuson-like phonons based on Ioffe-Regel limit criteria [74,75]. The first criterion compares the lifetime ($\tau$) of phonons with their period ($P$), stating that phonon modes with their $\tau$ smaller than $P$ cannot be treated as particles anymore and should be treated as diffusons. In this condition, phonon modes exhibit wavelike nature, which allows them to tunnel between close eigenstates and transport heat diffusively. Figure 6(a) shows the 3ph and 4ph rates, along with $P^{-1}$ (shown in the black-solid line). As seen, both 3ph and 4ph rates, even at the higher temperature of 2000K are smaller than $P^{-1}$ (equivalently $\tau > P$) showing that all the phonon modes are normal phonons. Similarly, the second criterion states that phonon modes with mean free path (MFP) smaller than minimum interatomic distance ($L_a$) become diffusons or diffuson-like phonons. As seen in Fig. 8(a), some phonon modes in the optical region (with frequency > 6 THz) have MFP < $L_a$ at 300 K and are diffuson-like phonons. The number of diffuson-like phonons increases with temperature and nearly one-third of the optical phonons become diffuson-like phonons at 2000 K. This explains the increase in diffuson thermal conductivity with temperature as more phonon modes become diffuson-like phonons at higher temperatures. Note that, in the Wigner formalism, all phonons are both particles and diffusons simultaneously at all temperatures. All phonons contribute to heat conduction through dual channels – particle and diffuson. As temperature increases, more phonons' diffuson nature weighs more than their particle nature.

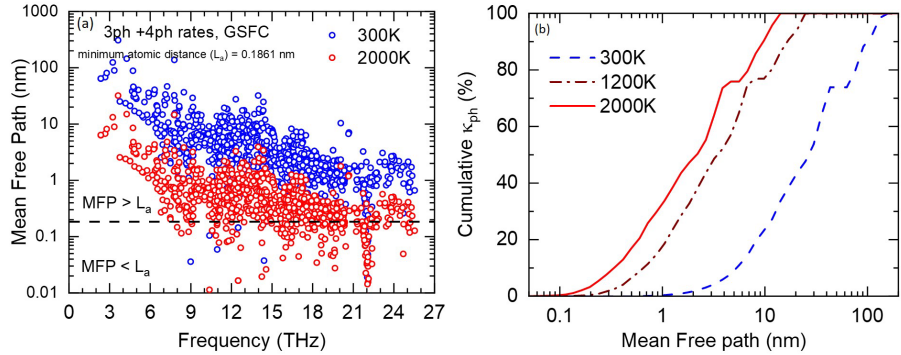

Figure 8:(a) mean free path of the phonon modes compared with minimum atomic spacing (b) cumulative phonon thermal conductivity in terms of mean free path lifetime.

Figure 8(b) shows the cumulative $\kappa_{ph}$ with respect to mean free path (MFP). At 300K, 80% of $\kappa_{ph}$ comes from phonon with MFP less than 50 nm. This value decreases to 5 nm as the temperature increases to 2000 K. As the experimental samples have grain size in the orders of μm, the κ reported in experimental studies are not suffered by grain boundary scattering.

**B. Radiation heat transfer**

To look into the reason behind the radiation heat transfer, we calculate the spectral radiative properties of $Al_2O_3$ using the Lorentz oscillator model, which uses the frequency and damping of IR active phonon modes at the Γ point. Based on symmetry analysis [76–78], the phonon modes on $Al_2O_3$ are $2A_{1g} + 2A_{1u} + 3A_{2g} + 2A_{2u} + 4E_u + 5E_g$. Among these modes, the $A_{2u}$ (extraordinary ray with the electric field vector parallel to the z-axis) and Eu (ordinary ray with the electric field vector perpendicular to the z-axis) species are IR-active modes, $A_{1g}$ and Eg are Raman-active modes, and $A_{2g}$ and $A_{1u}$ are spectroscopically inactive. The details on determining TO and LO branch indexes of IR active modes are discussed in Ref. [72,79].

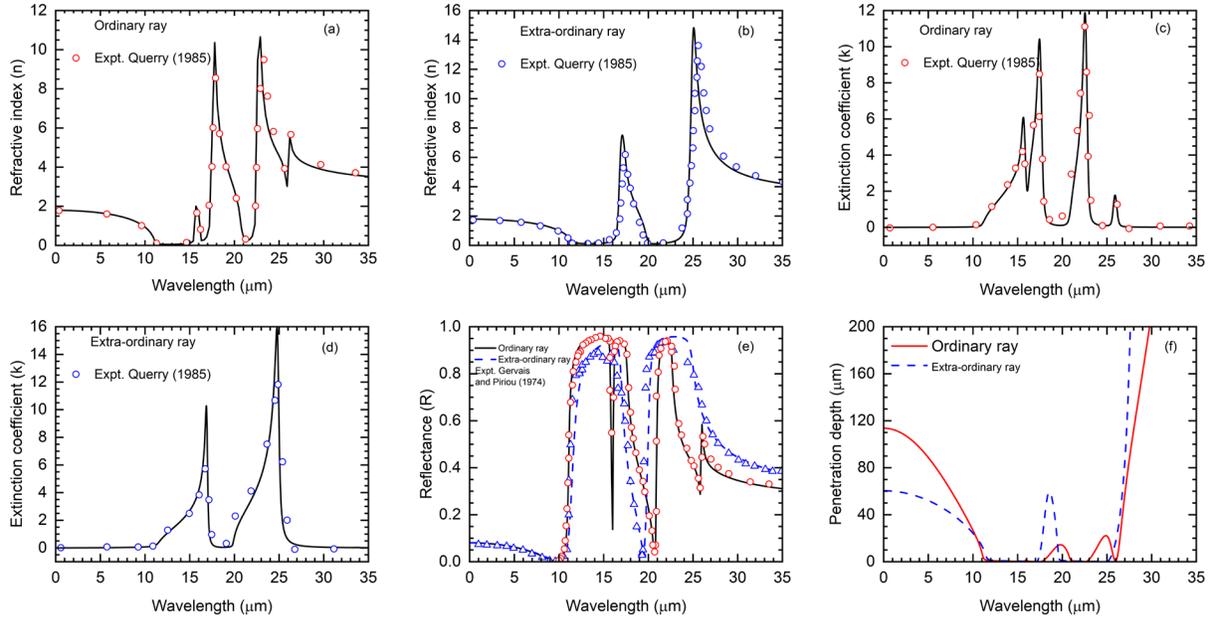

*Figure 9: Radiative thermal properties of Al$_2$O$_3$ at room temperature calculated in this work from first principles (curves) compared to experimental data (symbols). Refractive index (a) Ordinary ray (b) Extra-ordinary ray. Extinction coefficient (c) Ordinary ray (d) Extra-ordinary ray. (e) Reflectance. (f) Penetration depth.*

The radiative properties evaluated at room temperature are shown in Fig. 9. The spectral refractive index (*n*) and extinction coefficient (*k*) are shown in Fig. 9(a), 9(b), 9(c), and 9(d) respectively. The experimental data from Querry [80] are also shown for comparison, which shows a close agreement. The reflectance is calculated from the dielectric function or refractive index as:

$$R = \left| \frac{\sqrt{\epsilon}-1}{\sqrt{\epsilon}+1} \right|^2 = \frac{(1-n)^2 + k^2}{(1+n)^2 + k^2} \ . \qquad (13)$$

The spectral reflectance is shown in Fig. 9 (e), which matches well with the experimental data [48,71].

Extinction coefficient measures the attenuation of radiative waves inside the medium and is inversely correlated to photon MFP. It depends on the imaginary part of the dielectric function and is sensitive to the damping factor or phonon linewidth. Thus, temperature-dependent phonon linewidth is used to accurately calculate the extinction coefficient at higher temperatures. Note that the value for the extinction coefficient varies with the size of the material and was reported from 0.02 for bulk material [80] to 0.00008 for the thin film of 500 nm [81] at 1 μm wavelength. This results in significantly different κ$_{rad}$ on bulk materials and thin film. The present calculation is based on the bulk material and is compared with the experimental data reported for bulk material by Querry [80]. From the spectral extinction coefficient, we can see that the $Al_2O_3$ is nearly transparent in the near-IR range, with a penetration depth of around 110 μm for ordinary rays and 60 μm for extra-ordinary rays in this range (Fig. 9(f)). In the case of thin film, the photon might pass through the material instead of interacting with it.

The present calculation is based on the Rosseland model, which assumes the materials to be optically thick, where the photon gets absorbed in the medium and re-emitted and re-absorbed. In this scenario, the medium behaves as a participating medium and leads to the radiation thermal transport, as discussed in the previous papers [18,21,24,43–46]. The dominant radiation in the high-temperature region lies in the near-infrared range, as per Wein's law ($\lambda_{dom}T = 2898$ μm · K), for which the penetration depth is in the order of ~100 μm. Since the experimental sample are in mm or higher [18,19,71,80], the optically thick medium approximation used in our calculations is justified.

## C. Effect of Non-Analytical correction

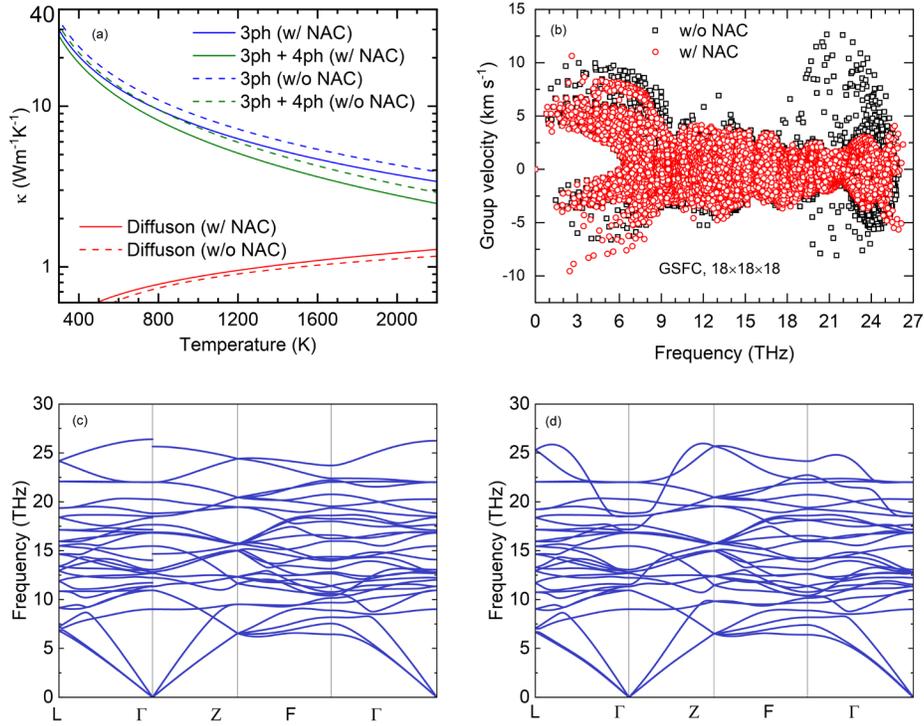

*Figure 10: Effect of Non-analytical correction (NAC) on (a) phonon thermal conductivity, (b) group velocities of phonon, and (c,d) phonon band dispersion.*

As shown in Fig. 10(a), the consideration of NAC decreases the $\kappa_{ph}$ moderately and increases $\kappa_{dif}$ slightly. The comparison is made for the calculation made with the ground-state force constant. NAC consideration decreases $\kappa_{ph}$ from 30.58, 5.95, and 2.90 Wm$^{-1}$K$^{-1}$ to 26.30, 5.09, and 2.48 Wm$^{-1}$K$^{-1}$ at the temperature of 300, 1200, and 2200 K respectively. Whereas $\kappa_{dif}$ changes from 0.48, 0.90, and 1.24 to 0.44, 0.95, and 1.28 Wm$^{-1}$K$^{-1}$ respectively. Overall, the $\kappa$ is overestimated if the NAC is not considered. This overestimation is due to the higher group velocities without NAC (Fig. 10(c)). The phonon scattering rates are not affected significantly. The change of group velocities is understandable since NAC causes TO-LO branch splitting, which flattens some branches in phonon dispersion as shown in Fig. 10 (c) and (d), and then reduces the group velocities.

## V. Conclusions

In conclusion, this work presents the accurate first-principles prediction of the thermal conductivity of $Al_2O_3$ from room temperature to near melting point (2200 K). The lattice thermal conductivity is found to be composed of contributions of phonon, diffuson, and radiation. The following conclusions are drawn. (1) Including all-temperature effects on phonon, diffuson, and radiation can reproduce the flatting and increasing trend of lattice thermal conductivity at high to ultra-high temperatures. (2) Phonon particle thermal conductivity decays approximately as $\sim T^{-1.14}$. Diffuson thermal conductivity increases roughly as $\sim T^{0.43}$. Radiation thermal conductivity increases as $\sim T^{2.51}$. (3) At room temperature, phonon, diffuson, and radiation contribute 98.7%, 1.3%, and zero, respectively. (4) At 2200 K, they contribute 61.2%, 19.7%, and 19.1%, respectively. (5) Four-phonon scattering is important at ultra-high temperature, decreasing the phonon thermal conductivity by a maximum of 24%. (6) The finite-temperature softening effects of harmonic and anharmonic force constants can increase the phonon thermal conductivity by a maximum of 36% at ultra-high temperatures. (7) The thermal conductivity from Green-Kubo MD agrees reasonably well with Wigner formalism, indicating that Green-Kubo MD captures phonons' both particle and wave natures. (8) The dominant phonon mean free path of $Al_2O_3$ is 50 nm and 5 nm at 300 K and 2200 K, respectively, indicating that it does not suffer from size effect for most experimental samples. (9) The photon penetration depth is about 100 nm, indicating that the ballistic effect of photon transport needs to be considered in the measurement of thermal conductivity of $Al_2O_3$ thin films when the film thickness is in the order of 100 nm at high temperatures. We hope this study has deepened the understanding of lattice thermal conductivity at ultra-high temperatures for complex crystals and will lead to more materials exploration for ultra-high temperature applications.

ACKNOWLEDGMENTS

This work is supported by the National Science Foundation (NSF) (award number: CBET 2212830). This work used Bridges-2 at Pittsburgh Supercomputing Center through allocation PHY220002 from the Advanced Cyberinfrastructure Coordination Ecosystem: Services & Support (ACCESS) program, which is supported by National Science Foundation grants #2138259, #2138286, #2138307, #2137603, and #2138296.The support and resources from the Center for High Performance Computing at the University of Utah are gratefully acknowledged.

Data availability

Source data are provided along with this paper. All other data that support the plots within this paper are available from the corresponding authors on reasonable request.

Code availability

The codes used in this study are available from the corresponding authors upon request.

Author contributions

T.F. conceived the idea and guided the project. J.T performed the simulations and wrote the original manuscript. J.T and T.F. both revised the manuscript.

Competing interests

The authors declare no competing interests.

Additional information

Correspondence and requests for materials should be addressed to T.F.